\documentclass{article}
\usepackage{graphicx}  
\usepackage{amsmath}   
\usepackage[compress]{cite}
\usepackage{amssymb}   
\usepackage{bm} 
\usepackage{dcolumn}
\usepackage{color}
\usepackage{mathrsfs}
\usepackage{amsfonts}
\usepackage{varioref}
\usepackage{textcomp}
\RequirePackage[colorlinks,citecolor=blue,urlcolor=magenta,linkcolor=blue]{hyperref}
\allowdisplaybreaks
\def\gr{general relativity}

\labelformat{section}{Section #1} 
\labelformat{subsection}{Section #1} 
\labelformat{subsubsection}{Section #1}
\labelformat{subsubsubsection}{Section #1}
\labelformat{equation}{Eq.~(#1)} 
\labelformat{figure}{Fig.~#1} 
\labelformat{subfigure}{Fig.~\thefigure#1} 
\labelformat{table}{Tab.~#1} 
\labelformat{appendix}{Appendix #1}
\title{Raychaudhuri equation with zero point length}
\author{ Sumanta Chakraborty \footnote{sumantac.physics@gmail.com}~$^{1}$,
Dawood Kothawala \footnote{dawood@physics.iitm.ac.in}~$^{2}$ and
Alessandro Pesci \footnote{pesci@bo.infn.it}~$^{3}$
\\
{$^{1}$\small{School of Physical Sciences, Indian Association for the Cultivation of Science, Kolkata-700032, India}}
\\
{$^{2}$\small{Department of Physics, Indian Institute of Technology Madras, Chennai-600040, India}}
\\
{$^{3}$\small{INFN Bologna, Via Irnerio 46, I-40126 Bologna, Italy}}
}
\begin{document}
  
\maketitle
\begin{abstract}
The Raychaudhuri equation for a geodesic congruence in the presence of a zero-point length has been investigated. This is directly related to the small-scale structure of spacetime and possibly captures some quantum gravity effects. The existence of such a minimum distance between spacetime events modifies the associated metric structure and hence the expansion as well as its rate of change deviates from standard expectations. This holds true for any kind of geodesic congruences, including time-like and null geodesics. Interestingly, this construction works with generic spacetime geometry without any need of invoking any particular symmetry. In particular, inclusion of a zero-point length results into a non-vanishing cross-sectional area for the geodesic congruences even in the coincidence limit, thus avoiding formation of caustics. This will have implications for both time-like and null geodesic congruences, which may lead to avoidance of singularity formation in the quantum spacetime.
\end{abstract}
\section{Introduction}\label{Sec:Intro}

Raychaudhuri equation governs the flow of geodesics in a given spacetime manifold and it has been the cornerstone in our understanding of formation of trapped surfaces and singularities (for a recent review, see \cite{Kar:2006ms}). Unlike the field equations, the Raychaudhuri equation has no connection a priori to the gravitational theory one is interested in, since it is purely of geometrical origin. It essentially determines the rate of change of area along a geodesic congruence, which gets connected to shear and rotation of the geodesic congruence and the component of Ricci tensor projected along the geodesics. Only when one tries to connect the Ricci tensor with matter energy momentum tensor, the gravitational field equations come into play. In Einstein gravity, with reasonable assumptions on the matter energy momentum tensor, the Raychaudhuri equation demonstrates that the geodesics will converge forming caustics (see also \cite{Iosifidis:2018diy}). This is broadly due to the attractive nature of gravity. In most of the situations these caustics do not lead to any spacetime singularities, but under certain circumstances they do, leading to formation of black hole or cosmological singularities. Removal of these curvature singularities has remained a puzzle for decades. In this work, we will present a novel approach where formation of caustics can be avoided which possibly will lead to avoidance of curvature singularities as well \cite{KotE,KotF,StaA}.  

It is generally believed that the quantum theory of gravity, as and when it comes into existence must take care of these curvature singularities. Since we do not have any consistent quantum theory of gravity yet in sight, one can not attack the problem of singularity removal head on, but can take a cue from various other attempts. The single most important fact that is common to all the candidate theories of quantum gravity is the existence of a zero-point length \cite{GarA,HosA}. We will incorporate this fact in the spacetime geometry by postulating that as two points on the manifold coincide, the geodesic distance between them does not vanish. As a consequence the classical metric $g_{ab}$ gets modified to an effective metric $q_{ab}$ (which we will call the qmetric). The qmetric provides a squared geodesic interval between two events $P$ and $p$ which approximates to that provided by $g_{ab}$ in the limit of large geodesic distances, while at the same time approaches a finite value different from zero in the coincidence limit, i.e., as $p\to P$ \cite{KotE,KotF,StaA}. Note that the above approach incorporates some relics of quantum gravity irrespective of any specific theory of gravitational interaction.

A distinguishing aspect of this approach corresponds to the fact that it can incorporate some generic quantum gravity effects, but is based on the comfort zone of standard differential geometry. This provides a useful and at the same time general tool in describing the small-scale quantum effects. Further it can also be argued that one can incorporate the qmetric to find out how far one can proceed concerning understanding of various quantum aspects of gravity, without embracing any specific theory of gravity. On this line, invoking qmetric in various situations of interest, one can arrive at intriguing results also supported by other candidate theories of quantum gravity. In particular, in the qmetric approach the spacetime becomes effectively two-dimensional while approaching the Planck's scale \cite{Pad05}. The dimensional reduction of spacetime near the Planck scale is well known and appears in a variety of other approaches to quantum gravity, which include string theory, causal dynamical triangulations, causal set theory and loop quantum gravity \cite{WitA,AmbB,CarlG,ModA}. Similar results stem from the small distance limit of Wheeler-DeWitt equation \cite{CarlC,CarlC2}, from the asymptotic safety of the theory \cite{ReuA,PerA,Lit} and to different other attempts based on existence of a minimum length \cite{HusA, GubiA, ModB, MazA, CouA, CouB} (see \cite{CarlH} for a review and further references on the issue).
 
From the structure of the qmetric, several hints have been extracted regarding a possible statistical nature of the field equations for gravity, with intimate connection to the entropy extremization principle \cite{Pad06}. This derivation is similar to the earlier results discussed in \cite{PadG,PadF} based on the macroscopic spacetime thermodynamics alone. Thus qmetric may provide a microscopic justification for thermodynamic behaviour of null surfaces \cite{Chakraborty:2015hna}. The key aspect to these observations is the realization that the cross-sectional areas of equi-geodesic hypersurfaces, remain finite in the coincidence limit.

Further investigation of this subject naturally calls for a description of Raychaudhuri equation in the spacetime geometry described by the qmetric. As we have described earlier, given the generality of the approach, possibly the result derived in the context of qmetric will not be restricted to any specific situation but applicable to various approaches to quantum gravity. As emphasized earlier an understanding of the Raychaudhuri equation in this context will be crucial to see if quantum effects can avoid singularity formation \cite{DadA}.

There are indeed several results concerning the Raychaudhuri equation in a certain quantum gravity setting, even if perhaps they are not as numerous and general as in the context of dimensional reduction. However for completeness we will discuss earlier results suggesting that after accounting for quantum effects singularity formation could be avoided or, at least, not inevitable.
For example, in \cite{DasA} an attempt to derive the quantum Raychaudhuri equation has been presented based on exploitation of pilot's wave formulation of quantum mechanics. However this assumes an assigned background geometry and hence ignores back-reaction effects of the matter. There are also results from the context of loop quantum cosmology exhibiting avoidance of singularity formation in the cosmological context \cite{BojA, AshD}. This is due to the repulsive terms of quantum origin in the Raychaudhuri equation, which takes over when approaching the would-be cosmological singularity \cite{Singh, Li, DasB}. Similar results exist in the context of space-like singularity formation during collapse of a massive star to a Schwarzschild black hole \cite{AshB}. Similar consideration of string theory, brane world models and theories beyond \gr\ provides mixed results \cite{DasB}. This is because the nature of additional terms in the Raychaudhuri equation in these contexts depend on the equation of state of the perfect fluid describing matter. Following this interesting body of works, our aim here is to derive the Raychaudhuri equation using the qmetric description and hence study the effect of zero point length on formation of caustics. We will present a unified formulations for the null as well as space/time-like geodesic congruences. Subsequently we will investigate the derived equations in the coincidence limit and hence explore the consequences of zero point length in focussing of geodesics.

The paper is organized as follows: We have provided a basic introduction to the qmetric and have discussed the effect of qmetric on the expansion of null as well as space-like and time-like geodesics in \ref{Sec:Expansion}. Taking a cue from this analysis we have discussed the Raychaudhuri equation and its coincidence limit in \ref{Ray_Eq_Sec}. Finally we conclude with a discussion on the results obtained. Some additional computations are presented in \ref{App_A}. 
\section{The qmetric and expansion of geodesics}\label{Sec:Expansion}

In a $D$ dimensional spacetime we consider a space-like, time-like or null congruence $\Gamma$ of affinely parameterized geodesics. In case $\Gamma$ is made out of space-like or time-like congruence of geodesics, we define the normalized tangent vectors $n_{a}$ to the geodesic curves as, $n_{a}=\{1/2\sqrt{\epsilon \sigma ^{2}(x,x')}\}\nabla _{a}\sigma ^{2}(x,x')$, where $\sigma ^{2}(x,x')$ is the geodesic distance between the spacetime points $x^{a}$ and $x'^{a}$ and $\epsilon=\pm 1$ for space-like/time-like geodesics. If $x^{a}$ denote the spacetime coordinates of a generic point on a geodesic $\gamma \in \Gamma$, then the qmetric $q_{ab}(x,x')$ at $x^{a}$ relative to the point $x'^{a}$ can be written as \cite{StaA}
\begin{eqnarray}\label{qab}
q_{ab}(x,x')=A \, g_{ab} + \epsilon \, \Big(\frac{1}{\alpha} - A\Big) \, n_{a}n_{b}~.
\end{eqnarray}
The above holds true if the two points $x^{a}$ and $x'^{a}$ are separated by space-like/time-like geodesics, i.e., when $\Gamma$ consists of space-like/time-like geodesics. For null geodesics, a slightly different structure is necessary. If $\ell^{a}$ is the tangent to a null geodesic $\gamma$, which is affinely parametrized by $\lambda$, it follows that $\ell^a = (d/d\lambda)^a$. For null geodesics one must introduce an additional structure through the null vector $k^a$, defined as $k^a \equiv 2u^a - \ell^a$, where $u^a$ is the four-velocity of any time-like observer at that spacetime point. The observer is chosen such that it satisfies the following conditions, $\ell_a V^a = -1$ and $g_{ab}k^a \ell^b = -2$. A priori these relations hold true at a fixed point on the null geodesic, but parallel transport helps one to define these relations all along the null geodesic $\gamma$. In terms of these two null vectors $\ell^{a}$ and $k^{a}$, one can express the qmetric for null separated events as \cite{PesN},
\begin{eqnarray}
q_{ab}(x,x')= \mathcal{A} \, g_{ab} -\Big(\frac{1}{\mathcal{\beta}} - \mathcal{A}\Big) \,\ell_{\left(a\right.}k_{\left.b\right)}~.
\end{eqnarray}
Here symmetrization comes with a factor of $(1/2)$. The structure of the qmetric for space-like/time-like geodesics depends heavily on the quantities $\alpha$ and $A$ respectively, both being functions of the squared geodesic distance $\sigma^2(x,x')$ between $x^{a}$ and $x'^{a}$ respectively. These two quantities are expressed as \cite{StaA}
\begin{eqnarray}\label{alpha_A}
\alpha = \frac{S}{\sigma^2 \, {S^\prime}^2};\qquad A = \frac{S}{\sigma^2} \,
  \bigg(\frac{\Delta}{\Delta _{S}}\bigg)^\frac{2}{D-1}~,
\end{eqnarray}  
where $S = S(\sigma^2)$ is the geodesic distance according to the qmetric, with $\lim _{x\rightarrow x'}S= \epsilon L_{0}^2$, which is finite. In the above expression `prime' denotes differentiation with respect to $\sigma^2$ and $\Delta$ is the Van Vleck determinant associated with the geodesic distance $\sigma ^{2}$ \cite{vVl, Mor, DeWA, DeWB} (see also \cite{Xen, VisA, PPV}), defined as,
\begin{eqnarray}\label{vanVleck}
\Delta(p, P)=- \frac{1}{\sqrt{g(x) \, g(x')}} \,\det \bigg[- (\nabla_a)_{x} \, (\nabla_b)_{x'} \,\frac{1}{2} \sigma^2(x,x')\bigg]~.
\end{eqnarray}  
Further we have introduced another quantity $\Delta_{S}$, which is defined as $\Delta_{S}(x,x') \equiv \Delta(x_{S},x')$, with $x^{a}_{S}$ being that point on $\gamma$ which has the property $\sigma^2(x_{S},x')=S(x,x')$. Along identical lines the quantities $\beta$ and $\mathcal{A}$ associated with the qmetric for the null geodesics are functions of the affine parameter $\lambda$ such that (see, e.g., \cite{PesN}),
\begin{eqnarray}\label{alpha_diamond}
\beta =\frac{1}{d\lambda_{S}/d\lambda};\qquad \mathcal{A}=\frac{\lambda_{S}^2}{\lambda^2} \,
\bigg(\frac{\Delta}{\Delta_{S}}\bigg)^{\frac{2}{D-2}}~.
\end{eqnarray}
Here $\lambda_{S}$ is the qmetric affine parametrization of $\gamma$ such that, $\lim _{x\rightarrow x'}\lambda_{S} \to L_{0}$ in the coincidence limit. Further, we also have $\Delta_{S}(x,x') = \Delta(x_{S},x')$, where $x_{S}^{a}$ is that point on $\gamma$ (on the same side of $x$) which satisfies the condition $\lambda(x_{S},x')=\lambda_{S}$. This completes the basic discussion regarding the qmetric for both space-like/time-like and null geodesics.

The main ingredient of Raychaudhuri equation is the expansion of the geodesics. For space-like/time-like geodesics the appropriate quantity to look for is the trace of the extrinsic curvature, which has the following expression, $K=\nabla_a n^a$. On the other hand, for affinely parametrized null geodesics, similar expression for the expansion reads $\theta = \nabla_a \ell^a$. Our main aim of this work is to discuss the expansion $K_q$ and its rate of change for space-like/time-like geodesics, as well as $\theta _{q}$ and its rate of change for null geodesics in the presence of a zero-point length. Ultimately we want to explore the behaviour of the resulting equations in the coincidence limit along $\gamma$.

In the non-null case we start from the results presented in \cite{KotG}. In which case for geodesic curves the trace of the extrinsic curvature associated with the qmetric reads
\begin{eqnarray}\label{rayqab2_32_4}
K_{q}=\sqrt{\alpha} \,\Big[K + (D-1) \, \frac{d}{d\sigma} \ln \sqrt{A}\Big]~,
\end{eqnarray}  
where $\sigma \equiv \sqrt{\epsilon \sigma^2}$. Here $K_q = \nabla_a^{(q)} n^a_{(q)}$, with $n_{a~(q)}=(1/2\sqrt{\epsilon S}) \nabla _{a}S$ is the tangent to $\gamma$ at $p$ according to the qmetric-affine parameterization. Further note that the covariant derivative is also defined with respect to the qmetric, leading to its own connection $\Gamma^{a}_{bc}(q)=\frac{1}{2} q^{ad}(-\nabla_d q_{bc} + 2 \nabla_{\left(b\right.}q_{\left.c\right)d}) + \Gamma^{a}_{bc}$, where $\Gamma^{a}_{bc}$ is the connection compatible with $g_{ab}$ \cite{KotG}. From \ref{alpha_A}, the parameter $\alpha$ can be rewritten as $(d\sqrt{\epsilon S}/d\sigma)^{-2}$. Using \ref{alpha_A} and \ref{rayqab2_32_4} we readily get
\begin{align}\label{Ray_K}
\bigg(\frac{dK}{d\sigma}\bigg)_{q}
=\frac{dK_{q}}{d\sqrt{\epsilon S}}=\alpha \, \frac{dK}{d\sigma}+(D-1) \, \alpha \, \frac{d^2\ln \sqrt{A}}{d\sigma^2}  
+\frac{1}{2} \, \frac{d\alpha}{d\sigma} \,\bigg[K+(D-1) \, \frac{d\ln \sqrt{A}}{d\sigma} \bigg]~,
\end{align}
which coincides with the expression reported in \cite{KotG} for the rate of change of expansion of congruences of space-like/time-like equi-geodesic curves associated with the qmetric. In the null case, on the other hand, the expansion $\theta_{q}$ associated with the qmetric takes the following form \cite{PesN,PesM}
\begin{align}\label{rayqab2_31_3}
\theta_{q}&=\nabla_a^{(q)} \, \ell^a_{(q)}=\left(\frac{d\lambda}{d{\lambda_{S}}}\right) \, \theta
+\frac{1}{2} (D-2) \, \frac{d\lambda}{d{\lambda_{S}}} \,\frac{d\ln \mathcal{A}}{d\lambda} 
=\beta \,\Big[\theta + (D-2) \, \frac{d\ln \sqrt{\mathcal{A}}}{d\lambda} \Big]~.
\end{align}
Here, $\nabla _{a}^{(q)}$ is the qmetric covariant derivative, which has been introduced after \ref{rayqab2_32_4} and $l^a_{(q)} = (d/d\lambda_{S})^a$ is the tangent to the null geodesics with qmetric-affine parameterization $\lambda_{S}$. Using the explicit expressions for the quantity $\mathcal{A}$ from \ref{alpha_diamond} in terms of the associated Van-Vleck determinant, we finally obtain,
\begin{align}\label{rayqab2_31_6}
\bigg(\frac{d\theta}{d\lambda}\bigg)_{q}&=\frac{d\theta_q}{d\lambda_{S}} 
=\beta \, \frac{d\theta}{d\lambda_{S}} +(D-2) \, \beta \, \frac{d}{d\lambda_{S}} \, \frac{d}{d\lambda} \ln \sqrt{\mathcal{A}} 
+\frac{d\beta}{d\lambda_{S}} \,\bigg[\theta + (D-2) \, \frac{d}{d\lambda} \ln \sqrt{\mathcal{A}}\bigg]
\nonumber 
\\
&=\beta^{2} \, \frac{d\theta}{d\lambda} +(D-2) \, \beta^{2} \,\frac{d^2}{d\lambda^2} \ln \sqrt{\mathcal{A}} 
+\frac{1}{2} \, \frac{d(\beta^{2})}{d\lambda} \,\bigg[\theta +(D-2) \, \frac{d}{d\lambda} \ln \sqrt{\mathcal{A}}\bigg]~.
\end{align}
This yields the rate of change of the expansion of the null generators along the null geodesic in the context of qmetric. It is interesting to note that the equations, namely \ref{Ray_K} and \ref{rayqab2_31_6} for space-like/time-like and null geodesics can be transformed from one to the other. This is achieved through the following replacement, namely, $\{(D-2), \beta^{2}, \mathcal{A}\} \leftrightarrow \{(D-1), \alpha, A\}$, or in other words, $\{(D-2), \lambda, \lambda_{S}\}\leftrightarrow (D-1), \sigma, \sqrt{\epsilon S}\}$. Note that, so far we have not used the explicit expressions for the quantity $A$ (or, $\mathcal{A}$). Use of which along with some expression for the extrinsic curvature in terms of Van Vleck determinant enables us to provide alternative, but simpler expressions of the Raychaudhuri equation in the presence of zero point length, which will be useful while considering the coincidence limit. 

For this purpose, we start with the following expression for the extrinsic curvature in terms of the Van Vleck determinant, namely, $K=\{(D-1)/\sigma\}-(d/d\sigma)\ln \Delta$. Inserting this expression in \ref{Ray_K} and using the expression for $A$ from \ref{alpha_A}, we obtain (for a derivation see \ref{App_A}),
\begin{align}\label{Ray_K_Mod}
\bigg(\frac{dK}{d\sigma}\bigg)_{q}&=-\frac{D-1}{\left(\sqrt{\epsilon S}\right)^{2}}-\frac{d^{2}\ln \Delta _{S}}{d\sqrt{\epsilon S}^{2}}~.
\end{align}
Thus we can relate the rate of expansion of space-like/time-like geodesics in the presence of zero point length with the modified geodesic distance and modified Van~Vleck determinant associated with the qmetric. It is possible to write down a similar expression for the rate of expansion of null geodesics as well. This requires use of the following expression for the expansion $\theta$ of null geodesics, such that, $\theta=(D-2)/\lambda-(d/d\lambda)\ln \Delta$. Use of this expression along with that for $\mathcal{A}$ as in \ref{alpha_diamond}, casts \ref{rayqab2_31_6} to the following form (see \ref{App_A} for derivation),
\begin{align}\label{Ray_Null_Mod}
\bigg(\frac{d\theta}{d\lambda}\bigg)_{q}&=-\frac{D-2}{\lambda_{S}^{2}}-\frac{d^{2}\ln \Delta _{S}}{d\lambda_{S}^{2}}~.
\end{align}
This provides the simpler form of the rate of change of expansion for null geodesics in the presence of zero point length. We would like to emphasize that, following our expectations, the rate of change of expansion for the space-like/time-like and the null case can be derived from one another through the following mapping: $\{(D-1), \sqrt{\epsilon S}\} \leftrightarrow \{(D-2), \lambda_{S}\}$. This completes our discussion regarding derivation of the rate of change of expansion for qmetric, inheriting zero point length, starting from the original classical spacetime, characterized by the metric $g_{ab}$ or the geodesic distance $\sigma^{2}$. We will now try to understand the coincidence limit, i.e., as the geodesics starts to converge. In particular, we would like to see whether the convergence of geodesics can be avoided in the present premise.  
\section{Coincidence Limit: Finiteness of Raychaudhuri Equation}\label{Ray_Eq_Sec}

In this section we will first write down the Raychaudhuri equation associated with geodesic observers for both space-like and null hypersurfaces and then shall discuss the coincidence limit of the Raychaudhuri equation and argue about finiteness of the same. This may have interesting implications for singularity structure in the presence of zero point length. First of all the Raychaudhuri equation associated with the expansion of time-like geodesics without the zero point length reads,
\begin{align}\label{Ray_K_01}
\bigg(\frac{dK}{d\sigma}\bigg)=-\frac{1}{D-1}K^2-\sigma_{ab}\sigma^{ab} -R_{ab}n^{a}n^{b}~,
\end{align}
where, $\sigma_{ab}=K_{ab}-\{1/(D-1)\}Kh_{ab}$ is traceless as $h_{ab}=g_{ab}+n_{a}n_{b}$ is the induced metric on the equi-geodesic surface, and $K_{ab} =h^{c}_{a}\nabla_{c}n_{b}=\nabla_{a}n_{b}$ (since $n_{a}$ satisfies geodesic equation). The twist $\omega_{ab}$ is absent in the above expression due to hypersurface orthogonality of the vectors $n_{a}$, tangent to the geodesic. As evident from the expansion of $K$, both $dK/d\sigma$ and $K^{2}$ diverges in the coincidence limit and hence in this limit the Raychaudhuri equation, presented above, becomes ill-defined. 

Even though the extrinsic curvature $K$ of the equi-geodesic surfaces scale as $(1/\sigma)$, the quantity $\sigma_{ab}\sigma^{ab}+R_{ab}n^{a}n^{b}$ is finite in the coincidence limit and takes the value,
\begin{align}
\lim _{\sigma \rightarrow 0}\left(\sigma_{ab}\sigma^{ab}+R_{ab}n^{a}n^{b}\right)
&=-\bigg(\frac{dK}{d\sigma}\bigg)-\frac{1}{D-1}K^2
\nonumber
\\
&=-\frac{1}{D-1}\left(\frac{D-1}{\sigma}-\frac{\sigma}{3}\mathcal{F}\right)^2-\bigg(-\frac{D-1}{\sigma^{2}}-\frac{1}{3}\mathcal{F}\bigg)=\mathcal{F}~,
\end{align}
where $\mathcal{F}\equiv R_{ab}n^{a}n^{b}$. Thus a part of the Raychaudhuri equation remains finite in the coincidence limit, while overall both the sides of the Raychaudhuri equation diverge. This signifies the formation of caustics as the geodesics meet at a certain point. 

The above conclusion was derived from general relativistic consideration. However, incorporation of a zero-point length in the spacetime will presumably prohibit formation of such caustics. Thus it will be interesting to ask what happens to the Raychaudhuri equation in the coincidence limit from the qmetric perspective, in particular can the associated geodesics form caustics? To answer that, we can immediately express the Raychaudhuri equation by appropriately generalizing \ref{Ray_K_01},  presented in the context of qmetric as,
\begin{align}\label{Ray_K_02}
-\frac{D-1}{\left(\sqrt{\epsilon S}\right)^{2}}-\frac{d^{2}\ln \Delta _{S}}{d\sqrt{\epsilon S}^{2}}
=\left(\frac{dK}{d\sigma}\right)_{q}
=-\frac{1}{D-1}K_q^2-\sigma_{ab}^{(q)}\sigma^{ab}_{(q)} -R^{(q)}_{ab}n_{(q)}^{a}n_{(q)}^{b}~,
\end{align}
where \ref{Ray_K_Mod} has been used to relate $(dK/d\sigma)_{q}$, appearing on the left hand side of the Raychaudhuri equation, to the modified geodesic distance $S(\sigma^{2})$ and derivative with respect to the modified Van-Vleck determinant $\Delta_{\rm S}$. From this and further inspection of the formula for $K_q$ in \ref{App_A} (see \ref{rayqab2_35_2}), we see that we need to know the expression for Van Vleck determinant as well as its first and second derivatives to comment on formation of caustics in this case. We have to be careful, since there exist no general expression for the Van Vleck determinant, but only some expansion for small $\sigma$. It is certainly possible to carry over that expansion to qmetric as well (see \ref{App_A}), but these series cannot converge if the curvature at $x'$ blows up. Even if the curvature is finite at $x'$, still $\Delta_S$ can be diverging at point $x_S$ if geodesics emerging from $x'$ do have a focal point at $x_S$ (due the meaning of Van Vleck determinant as ratio of the actual density of geodesics and the density for flat spacetime, cf. \cite{VisA}). If $L_{0}$ is of the order of Planck's length and we are not too near to a singularity (safely away with distance $\sim \mathcal{O}(L_{0})$), we can be sure that no such focal points can appear before a distance $L_{0}$ from $x'$, and thus $\lim_{x\to x'}\Delta_S$ is finite. From the finiteness of the expansion of ${\Delta_S}$, we can also deduce that its first and second derivatives will be finite. Thus the result we will derive next, has general direct applicability towards formation of caustics, but is not immediately applicable at an already formed singularity. For the second derivative, the expansion yields
\begin{align}\label{VVD_q}
\frac{d^{2}\ln \Delta _{S}}{d\sqrt{\epsilon S}^{2}}
=\frac{\mathcal{F}}{3}+\frac{\dot{\mathcal{F}}}{2}\sqrt{\epsilon S}+\mathcal{O}(\epsilon S)~,
\end{align}
where, $\mathcal{F}=R_{ab}n^{a}n^{b}$ and `dot' denotes derivative with respect to the geodesic distance $\sigma$. As mentioned, the above quantity is finite in the coincidence limit, and is proportional to $\mathcal{F}$ to the leading order. Analogously, even if $K$ diverges in the coincidence limit, $K_{(q)}$ does not. This can be seen by using \ref{rayqab2_32_4} and expressions for $\alpha$ and $A$ from \ref{alpha_A}, such that for spacelike geodesics,
\begin{align}\label{New_Eq_01}
\lim _{x\rightarrow x'} K_{q}&=\lim _{x\rightarrow x'} \frac{\sqrt{S}}{\sigma (dS/d\sigma^{2})}\left[\frac{D-1}{\sigma}-\frac{d \ln \Delta}{d\sigma}+(D-1)\frac{d}{d\sigma}\ln \left\{\frac{\sqrt{S}}{\sigma}\left(\frac{\Delta}{\Delta_{S}}\right)^{1/(D-1)} \right\}\right]
\nonumber
\\
&=\lim _{x\rightarrow x'} \frac{\sqrt{S}}{\sigma (dS/d\sigma^{2})}\left[\frac{D-1}{\sigma}-\frac{d \ln \Delta}{d\sigma}
-\frac{(D-1)}{\sigma}+\frac{(D-1)}{2\sqrt{S}}\frac{1}{\sqrt{S}}\frac{dS}{d\sigma}
+\frac{d \ln \Delta}{d\sigma}-\frac{d \ln \Delta_{S}}{d\sigma}\right]
\nonumber
\\
&=\lim _{x\rightarrow x'} \frac{\sqrt{S}}{\sigma (dS/d\sigma^{2})}\left[\frac{2(D-1)\sigma}{2S}\frac{dS}{d\sigma^{2}}
-\frac{d \ln \Delta_{S}}{d\sqrt{S}}\times \frac{2\sigma}{2\sqrt{S}}\frac{dS}{d\sigma ^{2}}\right]
\nonumber
\\
&=\lim _{x\rightarrow x'} \left[\frac{(D-1)}{\sqrt{S}}-\frac{d \ln \Delta_{S}}{d\sqrt{S}}\right]
\nonumber
\\
&=\left(\frac{D-1}{L_{0}}\right)-\frac{\mathcal{F}}{3}L_{0}-\frac{\dot{\mathcal{F}}}{4}L_{0}^{2}+\mathcal{O}(L_{0}^{3})~.
\end{align}
Here we have used the fact that, $K=(D-1)/\sigma-(d/d\sigma)\ln \Delta$, as well as \ref{VVDq} in \ref{App_A}. A similar analysis can be performed for timelike geodesics as well, yielding an identical result. It turns out that alike the extrinsic curvature for the qmetric, its rate of change, i.e., $(dK/d\sigma)_{q}$ is finite as well in the coincidence limit. This can be seen from \ref{Ray_K_Mod}, leading to the following result
\begin{align}\label{New_Eq_02}
\lim _{x\rightarrow x'}\left(\frac{dK}{d\sigma}\right)_{q}=-\frac{D-1}{L_{0}^{2}}-\frac{\mathcal{F}}{3}-\frac{\dot{\mathcal{F}}}{2}L_{0}~,
\end{align}
which is also finite. Here we have used the expansion of the term $\ln \Delta _{S}$ as presented in \ref{App_A}. Thus both $K_{q}$ and its rate of change along the geodesic are finite, as evident from \ref{New_Eq_01} and \ref{New_Eq_02}, while the respective expressions for general relativity are diverging. Thus presence of a zero point length has smoothened the divergent quantinties. Finally from \ref{Ray_K_02} we can immediately obtain the coincidence limit of $\sigma_{ab}\sigma^{ab}+R_{ab}n^{a}n^{b}$ associated with the qmetric, which yield,
\begin{align}\label{New_Eq_10}
\lim _{x\rightarrow x'}\left(\sigma_{ab}\sigma^{ab}+R_{ab}n^{a}n^{b}\right)_{q}
&=-\left(\frac{dK}{d\sigma}\right)_{q}-\frac{1}{D-1}K_{q}^{2}
\nonumber
\\
&=\frac{D-1}{L_{0}^{2}}+\frac{\mathcal{F}}{3}+\frac{\dot{\mathcal{F}}}{2}L_{0}
-\frac{1}{D-1}\left\{\left(\frac{D-1}{L_{0}}\right)-\frac{\mathcal{F}}{3}L_{0}-\frac{\dot{\mathcal{F}}}{4}L_{0}^{2}\right\}^{2}
\nonumber
\\
&=\mathcal{F}+\dot{\mathcal{F}}L_{0}+\textrm{terms~depending~on}\left(\mathcal{F}^{2},\ddot{\mathcal{F}}\right)\mathcal{O}(L_{0}^{2})~.
\end{align}
First of all, as expected, the above expression is finite, as in the case of $g_{ab}$, but more importantly inherits corrections over and above the general relativity result which are proportional to the zero point length $L_{0}$ and its higher powers. Thus both the left hand side and the right hand side of the Raychaudhuri equation for geodesics in qmetric are finite, in complete contrast with the corresponding situation with $g_{ab}$. This depicts another instance, where divergences in the qmetric manifest themselves in such a manner that geometric quantities derived from them are finite.

Another point must be emphasized in this context, expressions for quantities like, $\sigma_{ab}\sigma^{ab}$ as well as Ricci tensor associated with the qmetric are very difficult to determine in terms of geometrical quantities associated with $g_{ab}$. This has to do with the complicated non-local dependance of the qmetric on $g_{ab}$. Still, some components of the Ricci tensor associated with the qmetric can be presented in terms of geometrical quantities, which have highly non-trivial dependance on $g_{ab}$, see e.g., \cite{KotG,StaA}. Thus it would be interesting to use \ref{New_Eq_10} in order to check the consistency of any future computations connecting geometrical quantities associated with qmetric to that with $g_{ab}$.
 
This suggests that there will exist \emph{no} caustics and hence geodesic convergence can be avoided in the context of qmetric. It is tempting to comment on possible removal of curvature singularities as well in this context. This will happen in case of finiteness of the Van Vleck determinant $\Delta _{S}$ in the coincidence limit of collapsing matter world lines. We will have a look at this in next section.

A similar consideration applies to null geodesics as well, for which the Raychaudhuri equation associated with the background metric takes the familiar form,
\begin{align}\label{New_Eq_03}
\frac{d\theta}{d\lambda}=-\frac{1}{D-2}\theta^2-\sigma_{ab}\sigma^{ab} -R_{ab}\ell^{a}\ell^{b}~.
\end{align}
In this context as well, even though $\sigma_{ab}\sigma^{ab}$ and $R_{ab}\ell^{a}\ell^{b}$ are finite, the expansion squared and its rate of change along the null geodesic diverges in the coincidence limit. This again signals formation of caustics and convergence of null geodesics. The Raychaudhuri equation for null geodesics in the qmetric can be obtained by simply generalizing each geometrical quantities appearing in the above expression to their respective counterpart for qmetric. Further using \ref{Ray_Null_Mod}, the modified Raychaudhuri equation for qmetric becomes,
\begin{align}\label{Rayq_K_null}
-\frac{D-2}{\lambda_{\rm S}^{2}}-\frac{d^{2}\ln \Delta _{S}}{d\lambda_{S}^{2}}
=\left(\frac{d\theta}{d\lambda}\right)_{q}
=-\frac{1}{D-2}\theta_q^2-\sigma_{ab}^{(q)}\sigma^{ab}_{(q)} -R^{(q)}_{ab}\ell_{(q)}^{a}\ell_{(q)}^{b}~,
\end{align}
where $\sigma_{ab}=\theta_{ab}-\{1/(D-2)\}\theta~\chi_{ab}$ is the shear tensor associated with the null geodesics with $\chi_{ab}=g_{ab}+(1/2)(\ell_{a}k_{b}+\ell_{b}k_{a})$ being the induced metric on the equi-geodesic surface. In this case as well, in the coincidence limit the derivative of the Van Vleck determinant is given by \ref{VVD_q} with $\sqrt{\epsilon S}$ replaced by $\lambda_{S}$. Along with this the following results for coincidence limit of various geometrical quantities of interest can also be derived,
\begin{align}\label{New_Eq_04}
\lim _{x\rightarrow x'} \theta_{q}=\left(\frac{D-2}{L_{0}}\right)-\frac{\mathcal{F}}{3}L_{0}-\frac{\dot{\mathcal{F}}}{4}L_{0}^{2}+\mathcal{O}(L_{0}^{3})~;
\end{align}
and
\begin{align}\label{New_Eq_05}
\lim _{x\rightarrow x'}\left(\frac{d\theta}{d\lambda}\right)_{q}=-\left(\frac{D-2}{L_{0}^{2}}\right)-\frac{\mathcal{F}}{3}-\frac{\dot{\mathcal{F}}}{2}L_{0}+\mathcal{O}(L_{0}^{2})~,
\end{align}
where $\mathcal{F}$ is the null limit of $R_{ab}n^{a}n^{b}$, reading $R_{ab}\ell^{a}\ell^{b}$. These results can be derived by following the exact steps of \ref{New_Eq_01} and \ref{New_Eq_02}, keeping in mind that we are working with null geodesics. As evident from \ref{New_Eq_04} and \ref{New_Eq_05}, the quantities diverging in the coincidence limit for $g_{ab}$ are finite when their counterparts in the qmetric is considered. Thus we obtain, the coincidence limit of the geometrical quantity $\sigma_{ab}\sigma^{ab}+R_{ab}\ell^{a}\ell^{b}$ for the qmetric to be,
\begin{align}
\lim _{x\rightarrow x'}\left(\sigma_{ab}\sigma^{ab}+R_{ab}\ell^{a}\ell^{b}\right)_{q}
&=-\left(\frac{d\theta}{d\lambda}\right)_{q}-\frac{1}{D-2}\theta_{q}^{2}
\nonumber
\\
&=\frac{D-2}{L_{0}^{2}}+\frac{\mathcal{F}}{3}+\frac{\dot{\mathcal{F}}}{2}L_{0}
-\frac{1}{D-2}\left\{\left(\frac{D-2}{L_{0}}\right)-\frac{\mathcal{F}}{3}L_{0}-\frac{\dot{\mathcal{F}}}{4}L_{0}^{2}\right\}^{2}
\nonumber
\\
&=\mathcal{F}+\dot{\mathcal{F}}L_{0}+\textrm{terms~depending~on}\left(\mathcal{F}^{2},\ddot{\mathcal{F}}\right)\mathcal{O}(L_{0}^{2})~,
\end{align}
which is not only finite but also involve corrections proportional to various powers of the zero point length. Therefore, if we are not too close to already existing singularity (affine distance larger than orders of $L_0$) all of the previous discussion for spacelike/timelike case does apply also for the null case as well. Hence, in these circumstances, even in the context of null geodesics we have a finite coincidence limit for each term of the Raychaudhuri equation avoiding formation of caustics.

This is consistent with the result derived for time-like geodesics and to leading order is identical to $\mathcal{F}$. This provides yet another interpretation for the object $R_{ab}\ell^{a}\ell^{b}$, abundant in thermodynamic description of gravity \cite{Padmanabhan:2013nxa,Chakraborty:2014rga,Chakraborty:2015wma,Chakraborty:2015hna,Chakraborty:2017kob,Chakraborty:2018qew}. Thus our analysis explicitly demonstrates that the Raychaudhuri equation associated with qmetric remains finite in the coincidence limit, implying avoidance of caustics. This is because, there is always a residual length $L_{0}$ preventing the two geodesics from merging.  

Another interesting result in this context is non-vanishing of the cross-section of the geodesics in the coincidence limit. For time-like geodesics the effective cross-sectional region is a $(D-1)$-dimensional volume, while for null geodesics it is a $(D-2)$-dimensional area. In the context of qmetric both of them will be modified. It turns out that both the area and volume will be finite in the coincidence limit. In particular, the $(D-1)$-dimensional volume in the coincidence limit will behave as $d^{D-1} V_{q}=L_{0}^{D-1}(1/\Delta_{S}) (d\eta)^{D-1}$ and the $(D-2)$-dimensional surface will behave as $d^{D-2}A_{q}=L_{0}^{D-2}(1/\Delta_{S}) (d\eta)^{D-2}$. Here $(d\eta)^{D-1}$ (or, $(d\eta)^{D-2}$) is the angular contribution from the volume (or, area) of the respective region in coincidence limit (for details, see \cite{Pad06,PesN}). The finiteness of both these results are consistent with our findings from the Raychaudhuri equation for the qmetric. Since the fact that geodesics do not form caustics, as the coincidence limit is taken, ensures that the transverse area/volume normal to the geodesics must also remain finite. This provides yet another demonstration of the correctness of the result presented above.  
\section{Discussions and Concluding Remarks}

One of the key mathematical structures of a Lorentzian manifold is its causal structure, and {\it global} properties of this causal structure are crucial in understanding classical solutions of general relativity in the strong gravity regime. This is best demonstrated by the classical singularity theorems of Penrose and Hawking \cite{HawB}, the proofs of which crucially rely on the causal structure of the spacetime and some generic conditions on matter fields. However, what remains largely an unresolved issue is the behaviour of light cones, and the resultant causal structure of spacetime, at {\it small scales}. It is widely believed that quantum gravitational fluctuations would drastically affect the behaviour of light cones at small scales, thereby altering the causal connectedness of spacetime at very small scales. For example, in cosmology the BKL conjecture is effectively tied to the {\it closing up} of light cones near a space-like singularity. However, what happens to the light cones in a generic spacetime at an arbitrary event (not necessarily a singularity) remains largely unclear, although there have been analysis based on Raychaudhuri equation and stress tensor fluctuations \cite{carlip-prl,ford-etal1,ford-etal2,ford-etal3}. The analysis presented here is in a similar spirit, but attempts to go somewhat deeper, as we study the behaviour of light rays on a {\it quantum spacetime}, described by a qmetric, which admits a lower bound on geodesic intervals. This is perhaps the most minimalistic requirement that can be imposed on a quantum spacetime, supported by almost all known frameworks of quantum gravity. 

For completeness, let us briefly comment on a possible connection of this approach with string theory, in particular the notion of T-duality. For this purpose one should note that in most of the string theory models presence of extra dimensions are unavoidable and they must be compact to avoid detection at present day energy scales, leading to a compactification length scale in the theory. Therefore from the perspective of a four-dimensional observer the physics is bounded by the compactification length scale $R$ associated with the compactified extra dimensions. However, such a scenario must also respect T-duality inherited from underlying string theory, which cannot distinguish between $R$ and $\alpha'/R$, where $\alpha'$ is the string slope. This immediately suggests that from the perspective of a low energy observer, $\sqrt{\alpha'}$ acts as the minimum length scale from the perspective of a low energy observer. Thus in this manner one can motivate the existence of such a minimum length scale from the perspective of a higher dimensional string theory \cite{Spallucci:2005bm,Fontanini:2005ik}.

When generalised to null intervals \cite{PesN}, the qmetric provides new insights into the small scale behaviour of light cones emanating from an arbitrary event in spacetime. These insights strengthens further as we inspect the Raychaudhuri equation on the quantum spacetime, which is what has been attempted in the present work. Two key results emerge from this analysis: (i) existence of an upper bound on the expansions of null and time-like geodesics, and (ii) additional terms in the Raychaudhuri equation related to the Van~Vleck determinant associated with the modified geodesic interval. (For a result similar to (i), see \cite{carlip-prl}). As stressed in the derivation, these results hold true provided we are not too close to an already existing singularity. But what about if we have no singularity at start? Will zero-point length analysis foresee avoidance of singularity formation?  To investigate this, following \cite{PesM} we may consider a null shell, let us say a shell of photons, undergoing spherically symmetric collapse towards a spacetime point $C$. Our geodesics are now explicitly actual world lines of particles. Classically, a curvature singularity blatantly develops at $C$. This is because energy per unit transverse area diverges and the geodesics become incomplete \cite{HawB}. In the qmetric picture the situation is quite different. The energy density does not diverge as the van Vleck determinant $\Delta_S$ and then the area element remain finite in the coincidence limit. To see this, note that at coincidence, $\Delta_S$ is determined by a configuration in which no singularity is present, with the photons at points $x_S$ at affine distance $L_0$ from the point $C$, point in which everything is finite and regular. We can be sure thus that the points at $x_S$ are not focal points and then that $\Delta_S$ is finite. Thus, the null geodesics do not cease to exist after a finite affine parameter and one hopes that a singularity never develops. 

Hence, the most important implications of our analysis would be to study the structure of spacetime near a about-to-form space-like singularity, that is in a domain where time-like and null geodesics terminate, resulting in geodesic incompleteness, usually also accompanied by divergences in the curvature tensor components measured in some parallel propagated basis. Detailed quantitative predictions remain a challenging task. Indeed, it is worth emphasising here that our entire framework, based as it is on the structure of the qmetric, depends on the knowledge of the world function and the Van Vleck determinant. Exact expressions for these are not available even for the Schwarzschild geometry, while an approximate expansion in a covariant Taylor series would not be of much help at circumstances in which $\mathcal F$ is large.The essential complication we are hinting at can be conveyed by a simple consideration. We expect, on generic grounds, that the qmetric corrections would depend on the ratios $q_1=L_0^2/\sigma^2$ and $q_2=R L_0^2$, $R$ being a typical magnitude of the curvature tensor components. Away from a curvature singularity, we expect $q_1 \gg q_2$ in the coincidence limit. However, near a curvature singularity, $R$ itself might diverge as $1/\sigma^2$ (as happens for radial geodesics in Schwarzschild), thereby making $q_2 \sim q_1$. It is therefore impossible to find a domain in which any kind of Taylor expansion would be applicable. The only way forward seems to be to find a non-covariant expansion of the world function and the Van Vleck determinant in terms of some suitably chosen coordinates near the about-to-be singular region. This is currently being investigated.

It is worth noting, finally, that our derivation of the quantum Raychaudhuri equation does not hinge on any assigned particular symmetry of spacetime (like isotropy, for instance), and as such it refers to a completely generic geometry. This makes it applicable to arbitrary Lorentzian spacetimes, including the Lorentzian geometries arising as solutions to higher dimensional and/or higher curvature actions. Moreover, we have not made any assumptions regarding the nature of quantum fluctuations or of the matter stress-tensor that are responsible for distorting the causal structure of spacetime. Indeed, our results hold in the coincidence limit as long as geodesic intervals have a lower bound, and is insensitive to the exact form of the modified geodesic intervals (provided they satisfy certain smoothness conditions, see \cite{StaA}), which will anyway require a complete quantum gravitational analysis. In this sense, we expect our result concerning small scale behaviour of the Raychaudhuri equation on a quantum spacetime to be robust. It's implications for singularities and singularity theorems are under investigation. 
\section*{Acknowledgement}

Research of SC is supported by the INSPIRE Faculty Fellowship (Reg. No. DST/INSPIRE/04/2018/000893) from Department of Science and Technology, Government of India. The authors also thank T. Padmanabhan for his useful comments on an earlier version of this manuscript.
\appendix
\labelformat{section}{Appendix #1}
\labelformat{subsection}{Appendix #1}
\labelformat{subsubsection}{Appendix #1}
\section{Some Relevant Computations}\label{App_A}

In this appendix we will briefly describe some calculations relevant for the present work. First of all let us derive \ref{Ray_K_Mod} starting from the following expression for the extrinsic curvature in terms of the Van-Vleck determinant, namely, $K=\{(D-1)/\sigma\}-(d/d\sigma)\ln \Delta$. Inserting this expression in \ref{Ray_K} and using the expression for $A$ from \ref{alpha_A}, we obtain,
\begin{align}
\bigg(\frac{dK}{d\sigma}\bigg)_{q}&=\alpha \left\{-\frac{D-1}{\sigma^{2}}-\frac{d^{2}}{d\sigma ^{2}}\ln \Delta 
+(D-1) \frac{d^2}{d\sigma^2} \ln \left(\frac{\sqrt{\epsilon S}}{\sigma}\left(\frac{\Delta}{\Delta _{S}}\right)^\frac{1}{D-1}\right) \right\}
\nonumber
\\
&+\frac{1}{2} \, \frac{d\alpha}{d\sigma}
\left[\frac{D-1}{\sigma}-\frac{d}{d\sigma}\ln \Delta+(D-1) 
\frac{d}{d\sigma}\ln \left(\frac{\sqrt{\epsilon S}}{\sigma}\left(\frac{\Delta}{\Delta _{S}}\right)^\frac{1}{D-1}\right) \right]
\nonumber
\\
&=\alpha \left\{(D-1)\frac{d}{d\sigma}
\left(\frac{1}{\sqrt{\epsilon S}}\frac{d\sqrt{\epsilon S}}{d\sigma}\right)-\frac{d}{d\sigma}
\left(\frac{d\sqrt{\epsilon S}}{d\sigma}\frac{d\ln \Delta _{S}}{d\sqrt{\epsilon S}} \right)\right\}
-\left(\frac{d\sigma}{d\sqrt{\epsilon S}}\right)^{2}\frac{d^{2}\sqrt{\epsilon S}}{d\sigma^{2}}
\left(\frac{D-1}{\sqrt{\epsilon S}}-\frac{d\ln \Delta _{S}}{d\sqrt{\epsilon S}}\right)
\nonumber
\\
&=-\frac{D-1}{\epsilon S}\alpha \left(\frac{d\sqrt{\epsilon S}}{d\sigma}\right)^{2}
+\frac{\alpha}{\sqrt{\epsilon S}}(D-1)\frac{d^{2}\sqrt{\epsilon S}}{d\sigma^{2}}
-\alpha \frac{d\ln \Delta _{S}}{d\sqrt{\epsilon S}}\frac{d^{2}\sqrt{\epsilon S}}{d\sigma^{2}}
-\alpha\frac{d^{2}\ln \Delta _{S}}{d\sqrt{\epsilon S}^{2}}
\left(\frac{d\sqrt{\epsilon S}}{d\sigma}\right)^{2}
\nonumber
\\
&-\left(\frac{d\sigma}{d\sqrt{\epsilon S}}\right)^{2}\frac{d^{2}\sqrt{\epsilon S}}{d\sigma^{2}}
\left(\frac{D-1}{\sqrt{\epsilon S}}-\frac{d\ln \Delta _{S}}{d\sqrt{\epsilon S}}\right)
\nonumber
\\
&=-\frac{D-1}{\epsilon S}-\frac{d^{2}\ln \Delta _{S}}{d\sqrt{\epsilon S}^{2}}~.
\end{align}
In a similar fashion it is also possible to write down an expression for the rate of expansion of null geodesics as well, which is presented in \ref{Ray_Null_Mod}. The derivation requires use of the expression for $\theta$ and that for $\mathcal{A}$, which casts \ref{rayqab2_31_6} to the following form,
\begin{align}
\bigg(\frac{d\theta}{d\lambda}\bigg)_{q}&=\beta^{2} \left[-\frac{(D-2)}{\lambda^{2}}-\frac{d^{2}}{d\lambda ^{2}}\ln \Delta
+(D-2) \,\frac{d^2}{d\lambda^2}\ln \left(\frac{\lambda_{S}}{\lambda}
\bigg(\frac{\Delta}{\Delta_{S}}\bigg)^{\frac{1}{D-2}}\right)\right]
\nonumber
\\
&+\frac{1}{2} \, \frac{d(\beta^{2})}{d\lambda} \,\bigg[\frac{D-2}{\lambda}-\frac{d\ln \Delta}{d\lambda} 
+(D-2) \, \frac{d}{d\lambda} \ln \left(\frac{\lambda_{S}}{\lambda}
\bigg(\frac{\Delta}{\Delta_{S}}\bigg)^{\frac{1}{D-2}}\right)\bigg]
\nonumber
\\
&=\beta^{2} \left[(D-2)\frac{d}{d\lambda}\left(\frac{1}{\lambda_{S}}\frac{d\lambda_{S}}{d\lambda}\right)
-\frac{d}{d\lambda}\left(\frac{d\lambda _{S}}{d\lambda}\frac{d\ln \Delta _{S}}{d\lambda _{S}} \right)\right]
-\left(\frac{d\lambda}{d\lambda_{S}}\right)^{2}\frac{d^{2}\lambda_{S}}{d\lambda^{2}}
\left(\frac{D-2}{\lambda_{S}}-\frac{d\ln \Delta _{S}}{d\lambda_{S}}\right)
\nonumber
\\
&=-\frac{D-2}{\lambda_{S}^{2}}\beta^{2} \left(\frac{d\lambda_{S}}{d\lambda}\right)^{2}
+\frac{\beta^{2}}{\lambda_{S}}(D-2)\frac{d^{2}\lambda_{S}}{d\lambda^{2}}
-\beta ^{2} \frac{d\ln \Delta _{S}}{d\lambda_{S}}\frac{d^{2}\lambda_{S}}{d\lambda^{2}}
-\beta^{2}\frac{d^{2}\ln \Delta _{S}}{d\lambda_{S}^{2}}\left(\frac{d\lambda_{S}}{d\lambda}\right)^{2}
\nonumber
\\
&-\left(\frac{d\lambda}{d\lambda_{S}}\right)^{2}\frac{d^{2}\lambda_{S}}{d\lambda^{2}}\left(\frac{D-2}{\lambda_{S}}-\frac{d\ln \Delta _{S}}{d\lambda_{S}}\right)
\nonumber
\\
&=-\frac{D-2}{\lambda_{S}^{2}}-\frac{d^{2}\ln \Delta _{S}}{d\lambda_{S}^{2}}~.
\end{align}
These are the two expressions used in the main text. Note that these results can also be arrived at from a completely different perspective. We will illustrate that as well for completeness. Let us start from the expression of trace of extrinsic curvature for space-like/time-like geodesics, which we have described earlier. Substitution of this expression in \ref{rayqab2_32_4}, yields,
\begin{align}\label{rayqab2_35_2}
K_{q}&=\frac{d\sigma}{d\sqrt{\epsilon S}}\left[\frac{D-1}{\sigma}-\frac{d}{d\sigma} \ln \Delta
+(D-1)\frac{d}{d\sigma}\ln 
\left(\frac{\sqrt{\epsilon S}}{\sigma}\left(\frac{\Delta}{\Delta _{S}}
\right)^\frac{1}{D-1}\right)\right]
\nonumber
\\
&=\frac{d\sigma}{d\sqrt{\epsilon S}}\left[\frac{D-1}{\sqrt{\epsilon S}}
\frac{d\sqrt{\epsilon S}}{d\sigma}-\frac{d\sqrt{\epsilon S}}{d\sigma}\frac{d}{d\sqrt{\epsilon S}} 
\ln \Delta_{S}\right]
=\frac{D-1}{\sqrt{\epsilon S}}-\frac{d}{d\sqrt{\epsilon S}} \ln \Delta_{S}~.
\end{align}
Taking another derivative of this expression with respect to the modified geodesic distance $\sqrt{\epsilon S}$, we obtain $(dK/d\sigma)_{q}$. One can immediately verify that the resulting expression is identical to \ref{Ray_K_Mod}. Finally for null geodesics as well one can use the expression for expansion parameter $\theta$ for the classical spacetime, yielding the modified expansion parameter $\theta_{q}$ for qmetric, such that,
\begin{align}\label{theta_2}
\theta_{q}&=\frac{d\lambda}{d\lambda_{S}}\left[\frac{D-2}{\lambda}-\frac{d}{d\lambda} \ln \Delta
+(D-2)\frac{d}{d\lambda}\ln \left(\frac{\lambda_{S}}{\lambda}\left(\frac{\Delta}{\Delta _{S}}\right)^\frac{1}{D-2}\right)\right]
\nonumber
\\
&=\frac{d\lambda}{d\lambda_{S}}\left[\frac{D-2}{\lambda_{S}}\frac{d\lambda_{S}}{d\lambda}-\frac{d\lambda_{S}}{d\lambda}
\frac{d}{d\lambda_{S}} \ln \Delta_{S}\right]
=\frac{D-2}{\lambda_{S}}-\frac{d}{d\lambda_{S}} \ln \Delta_{S}~.
\end{align}
This expression, as one can easily verify will lead to \ref{Ray_Null_Mod} as a derivative with respect to $\lambda_{S}$ is taken. Note that in these (exact) expressions, any dependence of $(dK/d\sigma)_{q}$ or $(d\theta/d\lambda)_{q}$ on $\alpha$ and $A$ or on $\beta$ and $\mathcal{A}$ have been translated into a dependence on $\sqrt{\epsilon S}$ or $\lambda_{S}$ 
and the modified Van Vleck determinant $\Delta_{S}$. The modified Van Vleck determinant $\Delta _{S}$ can be expanded in a power series for small $l_S$, with $l_S \equiv \sqrt{\epsilon S}$ for space-like/time-like geodesics and  $l_S \equiv \lambda_S$ for null geodesics, with coefficients depending on the Riemann tensor of the classical spacetime $g_{ab}$. These expansions have been used while considering the coincidence limit and hence it is beneficial to point it out here,
\begin{align}\label{VVDq}
\Delta _{S}=1+\frac{\mathcal{F}(x')}{6} \, l_S^{2}
+\frac{\dot{\mathcal{F}}(x')}{12} \, l_S^{3}
+\mathcal{O}\left(l_S^4\right)~.
\end{align}
Here $\mathcal{F}=R_{ab}n^{a}n^{b}$ and $\dot{\mathcal{F}}=n^{a}\partial _{a}\mathcal{F}$ for space-like/time-like geodesics, while $\mathcal{F}=R_{ab}\ell^{a}\ell^{b}$ and $\dot{\mathcal{F}}=\ell^{a}\partial _{a}\mathcal{F}$ for null geodesics. We have used this expression in the main text. 


\begin{thebibliography}{00}

\bibitem{Kar:2006ms} 
  S.~Kar and S.~SenGupta,
  ``The Raychaudhuri equations: A Brief review,''
  Pramana {\bf 69}, 49 (2007)
  [gr-qc/0611123].

\bibitem{Iosifidis:2018diy} 
  D.~Iosifidis, C.~G.~Tsagas and A.~C.~Petkou,
  ``Raychaudhuri equation in spacetimes with torsion and nonmetricity,''
  Phys.\ Rev.\ D {\bf 98}, no. 10, 104037 (2018)
  [arXiv:1809.04992 [gr-qc]].

\bibitem{KotE}
D. Kothawala,
``Minimal length and small scale structure of spacetime'',
Phys. Rev. D {\bf 88} (2013) 104029,
arXiv:1307.5618.

\bibitem{KotF}
D. Kothawala, T. Padmanabhan,
``Grin of the Cheshire cat:
Entropy density of spacetime as a relic from quantum gravity'',
Phys. Rev. D {\bf 90} (2014) 124060,
arXiv:1405.4967.

\bibitem{StaA}
D. Jaffino Stargen, D. Kothawala,
``Small scale structure of spacetime: van Vleck determinant and equi-geodesic 
surfaces'',
Phys. Rev. D {\bf 92} (2015) 024046,
arXiv:1503.03793.

\bibitem{GarA}
L.J. Garay,
``Quantum gravity and minimum length'',
Int. J. Mod. Phys. A {\bf 10} (1995) 145,
gr-qc/9403008.

\bibitem{HosA}
S. Hossenfelder,
``Minimal length scale scenarios for quantum gravity'',
Liv. Rev. Rel. {\bf 16} (2013) 2,
arXiv:1203.6191.

\bibitem{Pad05}
T. Padmanabhan, S. Chakraborty, D. Kothawala,
``Spacetime with zero point length is two-dimensional at the Planck scale'',
Gen. Rel. Grav. {\bf 48} (2016) 55,
arXiv:1507.05669.

\bibitem{WitA}
J.J. Atick and E. Witten,
``The Hagedorn transition and the number of degrees of freedom 
of string theory'',
Nucl. Phys. B {\bf 310} (1988) 291.

\bibitem{AmbB}
J. Ambj{\o}rn, J. Jurkiewicz and  R. Loll, 	
``Spectral dimension of the universe'',
Phys.  Rev.  Lett.  {\bf 95}  (2005)  171301,
hep-th/0505113.

\bibitem{CarlG}
S. Carlip,
``Dimensional reduction in causal set gravity'',
Class. Quantum Grav. {\bf 32} (2015) 232001, 
arXiv:1506.08775.

\bibitem{ModA}
L. Modesto,
``Fractal structure of loop quantum gravity'', 
Class. Quant. Grav. 26 (2009) 242002, 
arXiv:0812.2214.

\bibitem{CarlC}
S. Carlip
``Spontaneous Dimensional Reduction in Short-Distance Quantum Gravity ?'',
in AIP Conf. Proc. {\bf 1196} (2009) 72,
arXiv:0909.3329.

\bibitem{CarlC2}
S. Carlip, 
``The small scale structure of spacetime'',
in
{\it Foundations of space and time}
, edited by J. Murugan, A. Weltman, and G. F. R.
Ellis, 
(Cambridge University Press, 2012), 
arXiv:1009.1136.

\bibitem{ReuA}
O. Lauscher and M. Reuter,
``Ultraviolet fixed point and generalized flow equation of quantum gravity'', 
Phys. Rev. D {\bf 65} (2002) 025013, 
hep-th/0108040.

\bibitem{PerA}
R. Percacci and D. Perini,
``Should we expect a fixed point for Newton's constant ?'',
Class. Quantum Grav. {\bf 21} (2004) 5035, 
hep-th/0401071.

\bibitem{Lit}
D.F. Litim, 
``On fixed points of quantum gravity'',
in AIP Conf. Proc. {\bf 841} (2006) 322, 
hep-th/0606044.

\bibitem{HusA} 	
V. Husain, S.S. Seahra, E.J. Webster,
``High energy modifications of blackbody radiation and dimensional reduction'',
Phys. Rev. D {\bf 88} (2013) 024014,
arXiv:1305.2814. 

\bibitem{GubiA} 	
G. Gubitosi, J. Magueijo,
``Reappraisal of a model for deformed special relativity'',
Class. Quantum Grav. {\bf 33} (2016) 115021,
arXiv:1512.03268.

\bibitem{ModB} 	
L. Modesto, P. Nicolini,
``Spectral dimension of a quantum universe'',
Phys. Rev. D {\bf 81} (2010) 104040,
arXiv:0912.0220. 

\bibitem{MazA}
M. Maziashvili, 	
``Quantum-gravitational running/reduction of space-time dimension'',
Int. J. Mod. Phys. D {\bf 18} (2009) 2209,
arXiv:0905.3612.

\bibitem{CouA}
D.N. Coumbe,
``Hypothesis on the nature of time'',
Phys. Rev. D {\bf 91} (2015) 124040,
arXiv:1502.04320. 

\bibitem{CouB}
D.N. Coumbe,
``Quantum gravity without vacuum dispersion'',
Int. J. Mod. Phys. D {\bf 26} (2017) 1750119,
arXiv:1512.02519.

\bibitem{CarlH}
S. Carlip,
``Dimension and dimensional reduction in quantum gravity'',
Class. Quantum Grav. {\bf 34} (2017) 193001,
arXiv:1705.05417.

\bibitem{Pad06}
T. Padmanabhan,
``Distribution function of the atoms of spacetime and the nature of gravity'',
Entropy {\bf 17} (2015) 7420,
arXiv:1508.06286.

\bibitem{PadG}
T. Padmanabhan,
``Dark energy and gravity'',
Gen. Rel. Grav. {\bf 40} (2008) 529,
arXiv:0705.2533.

\bibitem{PadF}
T. Padmanabhan, A. Paranjape,
``Entropy of null surfaces and dynamics of spacetime'',
Phys. Rev. D {\bf 75} (2007) 064004,
gr-qc/0701003.

\bibitem{Chakraborty:2015hna} 
  S.~Chakraborty and T.~Padmanabhan,
  ``Thermodynamical interpretation of the geometrical variables associated with null surfaces,''
  Phys.\ Rev.\ D {\bf 92}, no. 10, 104011 (2015)
  [arXiv:1508.04060 [gr-qc]].

\bibitem{DadA}
N. Dadhich,
``Singularity: Raychaudhuri equation once again'',
Pramana {\bf 69} (2007) 23,
gr-qc/0702095.

\bibitem{DasA}
S. Das,
``Quantum Raychaudhuri equation'',
Phys. Rev. D {\bf 89} (2014) 084068,
arXiv:1311.6539.

\bibitem{BojA}
M. Bojowald, 
``Absence of singularity in loop quantum cosmology'', 
Phys. Rev. Lett. {\bf 86}, (2001) 5227,
gr-qc/0102069.

\bibitem{AshD}
A. Ashtekar,
``Singularity resolution in loop quantum cosmology: A brief overview'',
J. Phys. Conf. Ser. {\bf 189} (2009) 012003,
arXiv:0812.4703.

\bibitem{Singh}
P. Singh
``Are loop quantum cosmos never singular?''
Class. Quantum Grav. {\bf 26} (2009) 125005,
arXiv:0901.2750. 

\bibitem{Li}
L.-F. Li, J.-Y. Zhu,
``Thermodynamics in loop quantum cosmology'',
Adv. High Energy Phys. {\bf 2009} (2009) 905705,
arXiv:0812.3544.

\bibitem{DasB}
D.J. Burger, S. Das, S.S. Haque, N. Moynihan, B. Underwood,
``Towards the Raychaudhuri equation beyond general relativity'',
Phys. Rev. D {\bf 98} (2018) 024006,
arXiv:1802.09499.

\bibitem{AshB}
A. Ashtekar, M. Bojowald,
``Quantum geometry and the Schwarzschild singularity'', 
Class. Quantum Grav. {\bf 23} (2006) 391,
gr-qc/0509075.

\bibitem{PesN}
A. Pesci,
``Quantum metric for null separated events and spacetime atoms'',
Class. Quantum Grav. {\bf 36} (2019) 075009,
arXiv:1812.01275.

\bibitem{vVl} 	
J.H. van Vleck,
``The correspondence principle in the statistical interpretation of
quantum mechanics'',
Proc. Nat. Acad. Sci. USA {\bf 14} (1928) 178.

\bibitem{Mor}
C. Morette,
``On the definition and approximation of Feynman's path integrals'',
Phys. Rev. {\bf 81} (1951) 848.

\bibitem{DeWA}
B.S. DeWitt, R.W. Brehme,
``Radiation damping in a gravitational field'',
Annals Phys. {\bf 9} (1960) 220.

\bibitem{DeWB}
B.S. DeWitt,
{\it The dynamical theory of groups and fields}
(Gordon and Breach, New York, 1965).

\bibitem{Xen}
S.M. Christensen,
``Vacuum expectation value of the stress tensor in an arbitrary curved 
background: The covariant point-separation method'',
Phys. Rev. D {\bf 14} (1976) 2490.

\bibitem{VisA}
M. Visser,
``van Vleck determinants: geodesic focussing and defocussing in Lorentzian
spacetimes'',
Phys. Rev. D {\bf 47} (1993) 2395,
hep-th/9303020.

\bibitem{PPV}
E. Poisson, A. Pound, I. Vega,
``The motion of point particles in curved spacetime'',
Liv. Rev. Rel. {\bf 14} (2011) 7,
arXiv:1102.0529.

\bibitem{KotG}
D. Kothawala,
``Intrinsic and extrinsic curvatures in Finsler{\it{esque}} spaces'',
Gen. Rel. Grav. {\bf 46} (2014) 1836,
arXiv:1406.2672.

\bibitem{PesM}
A. Pesci,
``Effective null Raychaudhuri equation'',
Particles {\bf 1} (2018) 230,
arXiv:1809.08007.

\bibitem{Padmanabhan:2013nxa} 
  T.~Padmanabhan,
  ``General Relativity from a Thermodynamic Perspective,''
  Gen.\ Rel.\ Grav.\  {\bf 46}, 1673 (2014)
  [arXiv:1312.3253 [gr-qc]].
  
\bibitem{Chakraborty:2014rga} 
  S.~Chakraborty and T.~Padmanabhan,
  ``Evolution of Spacetime arises due to the departure from Holographic Equipartition in all Lanczos-Lovelock Theories of Gravity,''
  Phys.\ Rev.\ D {\bf 90}, no. 12, 124017 (2014)
  [arXiv:1408.4679 [gr-qc]].
  
\bibitem{Chakraborty:2015wma} 
  S.~Chakraborty,
  ``Lanczos-Lovelock gravity from a thermodynamic perspective,''
  JHEP {\bf 1508}, 029 (2015)
  [arXiv:1505.07272 [gr-qc]].
  
\bibitem{Chakraborty:2018qew} 
  S.~Chakraborty and R.~Dey,
  ``Noether Current, Black Hole Entropy and Spacetime Torsion,''
  Phys.\ Lett.\ B {\bf 786}, 432 (2018)
  [arXiv:1806.05840 [gr-qc]].

\bibitem{Chakraborty:2017kob} 
  A.~Mishra, S.~Chakraborty, A.~Ghosh and S.~Sarkar,
  ``On the physical process first law for dynamical black holes,''
  JHEP {\bf 1809}, 034 (2018)
  [arXiv:1709.08925 [gr-qc]].

\bibitem{HawB}
S.W. Hawking and G.F.R. Ellis,
{\it The large scale structure of space-time}, 
(Cambridge University Press, Cambridge UK, 1973).

\bibitem{ford-etal1}
H. Vieira, L. Ford, V. Bezerra, ``Spacetime geometry fluctuations and geodesic deviation",
arXiv:1805.05264.
%
\bibitem{ford-etal2}
J. Borgman, L. Ford, ``The Effects of Stress Tensor Fluctuations upon Focusing", 
Phys.\ Rev.\ D {\bf 70}, 064032 (2003),
arXiv:gr-qc/0307043.
%
\bibitem{ford-etal3}
L. Ford, T. Roman, ``Minkowski Vacuum Stress Tensor Fluctuations", 
Phys.\ Rev.\ D {\bf 72}, 105010 (2005), 
arXiv:gr-qc/0506026.

%
\bibitem{Spallucci:2005bm} 
  E.~Spallucci and M.~Fontanini, ``Zero-point length, extra-dimensions and string T-duality,''
  gr-qc/0508076.
  
%
\bibitem{Fontanini:2005ik} 
  M.~Fontanini, E.~Spallucci and T.~Padmanabhan,``Zero-point length from string fluctuations,''
  Phys.\ Lett.\ B {\bf 633}, 627 (2006) [hep-th/0509090].
  
%
\bibitem{carlip-prl}
S. Carlip, R. A. Mosna, J. P. M. Pitelli
``Vacuum Fluctuations and the Small Scale Structure of Spacetime'',
Phys. Rev. Letts. {\bf 107} (2011) 021303,
arXiv:1103.5993

\end{thebibliography}
\end{document}